\documentstyle[12pt,aaspp4]{article}
 
\slugcomment{}
\lefthead{Grundahl, VandenBerg \& Andersen}
\righthead{Str{\:o}mgren photometry of M13}
 
\newcommand{\nnot}{{\sl Nordic Optical Telescope}\ }
\newcommand{\uvbybeta}{$uvby\beta$\ }
\newcommand{\mm}{$m_1$\ }
\newcommand{\cc}{$c_1$\ }
\newcommand{\cz}{$c_0$\ }

\newcommand{\byz}{$(b-y)_0$\ }
\newcommand{\uyz}{$(u-y)_0$\ }

\newcommand{\ch}{CH\ }
\newcommand{\cn}{CN\ }

\newcommand{\feh}{$[Fe/H]$\ }

\newcommand{\teff}{$T_{eff}$\ }
\newcommand{\hipparcos}{Hipparcos\ }
\newcommand{\stromgren}{Str\"{o}mgren\ }

\begin{document}
 

\title{\stromgren Photometry of Globular Clusters: The Distance and Age of
       M13, Evidence for Two Populations of Horizontal-Branch Stars\footnote
       {
       Based on observations made with the Nordic Optical Telescope,
       operated on the island of La Palma jointly by Denmark, Finland,
       Iceland, Norway, and Sweden, in the Spanish Observatorio del
       Roque de los Muchachos of the Instituto de Astrofisica de
       Canarias. }
       }
 
\author{Frank Grundahl,
\footnote{National Research Council, Herzberg Institute of Astrophysics,
       Dominion Astrophysical Observatory, 5071 West Saanich Road, 
       Victoria, British Columbia, V8X 4M6, Canada; 
       Electronic mail: Frank.Grundahl@hia.nrc.ca}
Don A. VandenBerg
\footnote{University of Victoria, Department of Physics \& Astronomy, 
       PO Box 3055, Victoria, British Columbia V8W 3P7, Canada;
       Electronic mail: davb@uvvm.uvic.ca}
and Michael I. Andersen
\footnote{Nordic Optical Telescope, Apartado 474,
       Santa Cruz de La Palma, La Palma, E-38700, Spain;
       Electronic mail: andersen@not.iac.es}
       }
 
 
\begin{abstract}
 
We present deep CCD photometry of the globular cluster M13 (NGC$\,$6205) 
in the \stromgren $uvby\beta$ system, and determine a foreground 
reddening of $E(b-y) = 0.015 \pm 0.01$ mag. From a fit to the 
main--sequence of metal--poor subdwarfs with \hipparcos parallaxes, we 
derive $(m-M)_0 = 14.38 \pm 0.10$  which implies an age near 12 Gyr 
assuming [Fe/H] $= -1.61$ and [$\alpha$/Fe] $= 0.3$.  
The distance-independent (\byz, \cz) diagram indicates that M13 and metal--poor 
field subdwarfs of similar metallicity must be coeval to within $\pm\,$1 Gyr.  
In addition, we find that, at any given $(b-y)_0$ color, there is a large 
spread in the \cz index for M13 red-giant branch (RGB) stars.  We suspect 
that this scatter, which extends at least as faint as the base of the RGB, 
is most likely due to star-to-star variations in the atmospheric abundances 
of the CNO elements.
We also note the existence of what appears to be two separate stellar
populations on the HB of M13. Among other possibilities, it could arise 
as the result of differences in the extent to which deep mixing occurs 
in the precursor red giants.  

\end{abstract}
 
\keywords{globular clusters: general --- globular clusters: individual (M13) ---
          stars: evolution --- stars: horizontal branch --- stars: Population II}

 
 
\section{Introduction}
 
 Despite a concerted effort by many researchers over the past few decades
 (see, e.g., the reviews by VandenBerg, Bolte \& Stetson 1996, \cite{stetson96},
 \cite{sarajedini97} and references therein), it has not been possible to
 reach a consensus on either the absolute or the relative ages of the
 Galactic globular clusters (GCs).  Much of the controversy in this field can
 be traced to the difficulty of determining reliable values for the basic
 parameters that enter into a determination of the cluster age such 
 as distance, reddening, overall metallicity and detailed abundance patterns.
 
 In this respect, the \stromgren $uvby\beta$ photometric system offers 
 unique advantages over broad-band photometry since it can provide
 precise estimates of  \teff, [Fe/H] and surface gravity for F and G 
 stars, as well as the reddening, on a star-by-star
 basis (see Schuster \& and Nissen 1989a and references therein).  Furthermore,
 databases of quite homogeneous \stromgren photometry exist for large samples
 of stars, both metal--poor and metal--rich (Schuster \& Nissen 1988; 
 \cite{olsen83,olsen84}).  Many of these stars have
 new, high-precision determinations of their parallaxes from the \hipparcos
 mission, making them particularly useful for distance determinations
 via the main--sequence fitting technique.  Moreover, since the
 \stromgren system includes a $u$ filter, which is entirely on the
 short wavelength side of the Balmer jump, it can be used to probe the
 temperatures and gravities of horizontal-branch (HB) stars, including those
 on extended blue HBs (\cite{deboer95}).  
 
 As is well known, considerable evidence has accumulated during the past
 $\sim 20$ years which indicates that red giants in globular clusters have
 mixed much deeper into their nuclear-burning interiors than canonical
 models predict --- see, e.g., Kraft et al. (1998), \cite{kraft97} and the reviews
 by \cite{dacosta97} and \cite{kraft94} for discussions of the extensive
 literature on this subject.  Indeed, in order to explain the observed
 abundance patterns of C, N, O, Na, Al, and Mg (also see \cite{pilachowski96};
 \cite{shetrone96}), it seems to be necessary for the mixing to penetrate
 so closely to the H-burning shell that some helium must also be dredged
 up into the surface layers during the evolution (e.g., \cite{langer97},
 Sweigart 1997a,b, and \cite{cavallo98}). As discussed
 in \cite{sweigart97b}, even small amounts of He mixing could
 have large effects on the HB and account for at least part of the
 second-parameter phenomenon (i.e., the diversity in HB morphology
 among GCs having very similar chemical compositions).
 
 
\section{Observations and Data Reduction}
 
 All of the M13 observations were obtained using the \nnot on La Palma,
 Canary Islands, and \stromgren \uvbybeta filters during the period from
 June 27 to July 02, 1995.  The weather was excellent during the observing
 run, giving seeing values between 0\farcs43 and 0\farcs8, with a mean value
 close to 0\farcs60.  The detector used was a SITe 1024 square chip providing 
 a total field of $3 \times 3$ arcminutes. All photometric reductions of the 
 cluster frames were carried out using DAOPHOT, ALLSTAR, ALLFRAME, and DAOGROW 
 (see Stetson 1987, 1990, 1994).  In each frame, approximately 20000 stars 
 were detected. For the standard stars the residuals in the transformation 
 from the instrumental to the standard system were: $V$: 0\fm004; $(b-y)$: 0\fm005; 
 \mm: 0\fm008; \cc: 0\fm010 and $\beta$: 0\fm008. For M13 the error in the zero-point 
 for each filter is estimated to be of order 0\fm01.  The data reduction procedures 
 will be described in more detail in a forthcoming paper.

\section{Cluster Parameters}
 
 Before attempting to determine the distance to M13, it is necessary
 to know the amount of interstellar absorption towards the cluster as well as
 the stellar iron content (preferably the detailed chemical abundance patterns).  
 We have used our \stromgren photometry and the calibrations of 
 Schuster \& Nissen (1989a, Eq. 1) to determine the foreground reddening.  
 A number of their calibrating stars were included as standards in our observing
 program to ensure that our photometric system would be close to theirs.
 The M13 stars used included turn--off and subgiants with 
 $V$--magnitudes between 17.5 and 20.5 ($\sim 500$ stars).
 In this way, we derive a reddening of $E(b-y) = 0\fm015 
 \pm 0.01$, or $E(B-V) = 0.021$, which is quite typical of the values found for 
 M13 (e.g,~\cite{harris96}).  
 
 In principle, the cluster metallicity can also be determined from \stromgren 
 photometry.  Using (again) the calibrations of \cite{schuster89a}, we derive a 
 mean metallicity of $\feh = -1.7$ for $\sim 500$ stars.  However, we 
 found some indication of a possible error in the transformation of our 
 instrumental $v$ magnitudes \footnote{No such trends were evident for the other 
 filters or indices.} to the standard system as a function of \mm and, since an 
 error of 0\fm01 in the \mm index corresponds to an error of $\approx 0.1$ dex 
 in the estimated metallicity, the [Fe/H] value that we have derived for M13 
 cannot be considered robust.  As a result, we have adopted [Fe/H] $= -1.6 \pm 
 0.2$ as a compromise of our photometric determination and recent spectroscopic 
 results (\cite{kraft98}, \cite{carretta97}).  Finally, we adopt [$\alpha/$Fe] $= 0.3$
 which seems to be appropriate for the halo population (\cite{carney96}).  There 
 are no measurements of this quantity in main-sequence or turnoff stars in M13, 
 but see \cite{kraft97} and \cite{pilachowski96a} for pertinent discussions.
 
\subsection{The Cluster Distance and Age}
 
 Ideally, the distance to M13 would be determined by a main-sequence fit
 of the C-M diagram for the unevolved cluster stars to
 a {\it large} sample of local subdwarfs having the {\it same} metallicity
 as the cluster and very precise parallaxes.  Unfortunately, the number of
 subdwarfs with [Fe/H] $\approx -1.6$ and with relative errors in their
 parallaxes at the few percent level remains small.  Potentially-useful
 subdwarf candidates were selected from the list of
 \cite{nissen97}, with metallicities derived from \stromgren photometry and
 parallaxes determined by \hipparcos.  
 As in the case of M13, the individual reddenings and
 metallicities of the subdwarfs were determined using the
 calibrations of \cite{schuster89a}.
 
 Following the usual procedure (see, e.g., \cite{vandenberg96}), the dereddened
 colors of the subdwarfs were individually adjusted to compensate for
 the (small) differences in their metallicities relative to that of
 M13  using the isochrones of \cite{vandenberg98}, as  
 transformed to the \stromgren system using \cite{kurucz92} color--temperature
 relations.  
 In the fit of the M13 $((u-y)_0, V)$ and $((u-y)_0, u)$ C-M diagrams 
 to the resultant ``mono--metallicity'' subdwarf sequence, we used stars with 
 $\sigma(\pi)/\pi\le 0.07$, where $\pi$ is the parallax, and photometric
 metallicities in the range $-1.69\le\,$[Fe/H]$\,\le-1.57$.  Consequently,
 the model-based color corrections are all quite small for these stars.
 Furthermore, by adopting such stringent constraints on the relative
 parallax error, the Lutz--Kelker corrections (\cite{lutz73,lutz79}) amount
 to only a few hundredths of a magnitude, at most, and possible systematic
 biases for the whole sample are negligible as well (see the discussion by
 Pont et al.~1998)
 \footnote{Two of the stars used in the main--sequence fit
 (HD 64090 and HD 188510) are listed by \cite{gratton97} and \cite{pont98}
 as spectroscopic binaries or radial--velocity variables.  However,  
 \cite{carney94} has 19 velocity measurements for HD 188510 over 8.2 years, 
 and they find no indication of radial--velocity variations ($rms. \approx 0.8\,$km/s), 
 while five out of six measurements of the velocity of HD 64090 by \cite{stryker85} 
 are in excellent agreement. \cite{stryker85} quotes a semiamplitude limit of $14\,$km/s
 for their sensitivity, their one discrepant velocity value differs from the 
 mean of the other five by $27.4\,$km/s.
 Moreover, regardless of which C-M diagram is used in the 
 subdwarf fit, these two stars show no sign of being ``overluminous'' with
 respect to the other two that we have used (HD 34328 and HD 25329) and the MS
 fiducial.  Therefore, if HD 64090 and/or HD 188510 truly are binaries, the
 secondary component(s) must be much fainter than the primaries.  }.  
 We applied Lutz--Kelker corrections to the subdwarfs using the formula  
 \footnote{Because of the high precision of the parallaxes for 
 the selected stars, the choice of $n$ has no significant impact on the derived 
 distance modulus.} given by \cite{hanson79} with $n=2$.  Figure 1 illustrates 
 the $((u-y)_0,V)$ diagram for M13 with the subdwarfs overplotted. 
 The main--sequence fit yields $(m-M)_V = 14.44$ and $(m-M)_u = 14.49$ with 
 an uncertainty of $\pm 0.10$ mag.  
 
\placefigure{grundahl_fig1}      
 
 Isochrones from \cite{vandenberg98} for [Fe/H]$\,= -1.61$, [$\alpha$/Fe]$\,= 0.3$,
 and ages of 10, 12, and 14 Gyr have also been superposed on the data
 in Fig.~1.  These indicate that the age of M13 is close to 12 Gyr.  
 The reduction in age from values near 14--16 Gyr
 that were commonly obtained a few years ago is due less to revisions in the
 cluster distance than to improvements in the stellar models.  Indeed, the
 M13 distance modulus derived here is within 0.1 mag of the values determined
 by e.g., \cite{richer86} and \cite{buckley92} in the pre--\hipparcos era.
 It is, instead, the use of improved opacities and an equation of state that
 treats non-ideal effects (e.g., \cite{proffitt93}), the adoption of a revised
 bolometric correction scale (see \cite{vandenberg97}), and the assumption of
 [$\alpha$/Fe] $= 0.3$, instead of a scaled--solar heavy--element mix,
 that is primarily responsible for the reduced GC ages.  It should be
 noted that the present models do not include the
 effects of helium diffusion, which can be expected to reduce the age at a
 given turnoff luminosity by $\approx 1$ Gyr (see, e.g., \cite{proffitt91}). 
 
\section{The $($\byz, \cz$)$ Diagram}
 
 Having extensive \stromgren photometry for M13 stars fainter than the 
 main--sequence turnoff allows us to determine whether the cluster differs
 significantly in age from field stars of similar metallicity. For F and G 
 stars, \cz $= $\,\cc $-\,\, 0.2 E(b-y)$ is a measure of the size of the 
 Balmer jump, which, in turn, is a measure of surface gravity and hence of 
 evolutionary state.  Therefore, the \cc index can be used to distinguish 
 between evolved and unevolved stars and to determine the ages of stars near 
 the turnoff (see, e.g., Fig.~7 in \cite{vandenberg85}) independently of any 
 knowledge of their distances.  This approach has been used by 
 \cite{schuster89b} to determine ages for a large sample of metal--poor field 
 stars. In Figure 2, we have overplotted the photometry for stars in their 
 catalogs having metallicities between $-1.4$ and $-1.9$ on the (\byz, \cz) 
 diagram for M13.  It is immediately obvious from this figure that the field 
 stars follow the cluster locus closely, indicating that the two stellar 
 populations must have very nearly the same age.  The maximum difference 
 between the turnoff \cz values is $\pm$0\fm025, which translates into an 
 age difference of $\sim \pm 1$ Gyr, which, we emphasize, is completely 
 independent of the adopted distance scale.  We cannot, of course, exclude 
 the possibility that some of the field stars on the lower main sequence are 
 older, only that none of the evolved stars has a greater age than M13.  
 
\placefigure{grundahl_fig2}         
 
 A surprising feature of the data plotted in Fig.~2 is that,
 at any given \byz color, the M13 giant-branch stars
 exhibit a large spread (0\fm1 -- 0\fm15) in the \cz index.  
 As we have no spectra for the observed stars, it is not
 clear at this point how these \cz variations should be interpreted.
 However, as noted in \S 1, it is well known that stars on the M13 RGB
 encompass large variations in the C, N, O, Mg, and Al abundances,
 most likely due to deep mixing during the post--main--sequence evolution.
 Star--to--star variations in these elements are the most likely cause of
 the \cz variations, in view of the fact that several investigations
 (\cite{bond80,twarog94,twarog95}) have demonstrated that stars with strong \ch
 and \cn bands can have significantly lower \cc indices than similar stars of
 normal \ch and \cn strength. Spectroscopic observations of the stars showing 
 extreme \cc values are needed to identify the elements responsible for the 
 variations. 

 If the large spread in the \cc observations {\it is\/} due to variations in
 the elemental abundances due to mixing, then Fig.~2 would suggest that 
 such mixing commences
 near the base of the red-giant branch.  One cannot help but speculate that
 a thermal instability in the H-burning shell might be at least partly
 responsible for the observed scatter since \cite{vonrudloff88} have shown
 that the greatest potential for such an instability occurs just as a star
 begins to ascend the giant branch.  This possibility should be investigated
 further.

\section{The Horizontal Branch}
 
 We now turn to a brief discussion of the properties of the observed horizontal
 branch.  In Fig.~1, a ZAHB locus (VandenBerg et al.~1998), which
 represents the extension of the isochrones to the core He-burning phase,
 has been superposed on the data on the assumption of
 the adopted distance modulus and reddening.  It is apparent that the
 coolest of the HB stars are reasonably well-matched by the model ZAHB.
 However, at \uyz$ < 0.95$, the HB stars appear to be $\sim 0\fm4$ ``overluminous''
 compared to the theoretical locus.  We note that this
 ``anomaly'' has also been seen by Dorman et al.~(1998, private communication)
 in their {\it HST} observations of M13 (using the $F336W$ and $F555W$ filters).
 In fact, a preliminary analysis of \stromgren data that we have in 
 hand for NGC$\,$288 and NGC$\,$6752,
 which both possess blue HBs, show the same unexpected morphology.  This
 cannot be explained in terms of evolution away from a ZAHB since
 the evolution is slowest near the zero-age locus and most stars should,
 therefore, be found adjacent to it.  Unless the discrepancy is due to
 a problem with the color--$T_{\rm eff}$ relations derived from model 
 atmospheres, which seems unlikely, a different ZAHB must apply to these stars;
 i.e., M13 (as well as NGC$\,$288 and NGC$\,$6752) must possess at least
 two distinct HB populations.
 
 At first sight, these observations appear to be in remarkable agreement with
 the recent predictions by Sweigart (1997a,b),
 who has demonstrated that, if red-giants undergo sufficiently
 deep mixing as to modify their envelope helium abundances appreciably, then
 their decendants on the HB should be bluer and more luminous than
 predicted by canonical models.  In fact, there is some indication in his
 calculations (compare, e.g., Figs.~7 and 9 in Sweigart 1997b) that, for
 the hottest HB stars, the He-mixed models and those based on standard
 assumptions, will nearly coincide.  Thus, the near match of the canonical ZAHB
 with the lower-bound of the M13 stars fainter than $V \sim 17.7$ (see Fig.~1)
 can be understood.  On the other hand, \cite{rood98} have found that
 there is no apparent discrepancy between canonical HB models and M13's HB
 when comparisons between the two are made on C-M diagrams involving bands
 that are bluer than the {\it HST} F336W filter (which is close to the 
 \stromgren $u$).  So, at the present moment, it is not clear how this will 
 be resolved.
 
 
 Finally, we note that in a $(u-y, u)$ diagram both the ZAHB appropriate to the 
 coolest BHB stars
 and the locus of subgiant stars are very nearly horizontal and very well 
 defined; thus it is easy to measure the $u$ luminosity difference ($\Delta u$) 
 to high accuracy. From the isochrones of \cite{vandenberg98} we estimate that 
 $\Delta u$ changes at a rate of $\approx 0.065$ mag/Gyr. Given the high precision 
 available here we estimate that this method will provide the possibility to 
 measure the age difference between clusters of similar metallicity to better 
 than 0.5$\,$Gyr.
 
\section{Summary}
 
 The capability of \stromgren photometry to reveal important new features of
 GC C-M diagrams is undeniable.  In this investigation of M13, we have found
 that:
 1) The reddening of M13 is $E(b-y) = 0.015 \pm 0.01$, which corresponds
       to $E(B-V) = 0.021 \pm 0.014$.
 2) The cluster distance is $(m-M)_0 = 14.38 \pm 0.10$, based on C-M diagram
       fits to local subdwarfs with \hipparcos parallaxes.  This implies an
       age of $12\pm 1.5$ Gyr according to recent theoretical isochrones
       for [Fe/H] $= -1.61$ and [$\alpha$/Fe] $= 0.3$ (but which do not take
       He diffusion into account).  
 3) There is a large spread in \cc for stars from the base to
       the tip of the RGB, which is probably indicative of star-to-star
       variations in the CNO abundances. 
 4)  Field stars with metallicities similar to that of M13 are found to be
       coeval with M13 to within $\approx \pm 1$ Gyr.
 5)  The HB shows evidence for two distinct populations, which may be
       related to the observed scatter in \cc on the RGB.  It is also possible
       that we are seeing the signature of He mixing in upper RGB stars,
       as discussed by Sweigart (1997a,b).
 
\acknowledgements

We thank M\'arcio Catelan, Roger Bell, Ben Dorman, Poul Erik Nissen and James E. Hesser
for helpful remarks. Peter B. Stetson is thanked for letting us use his 
excellent photometry software.  FG gratefully acknowledges financial 
support from the Danish Natural Sciences Research Council.  He also 
acknowledges the hospitality offered by the National Research Council 
of Canada for making his stay at the Dominion Astrophysical Observatory 
possible.  DAV is grateful for the support of an Operating Grant from 
the Natural Sciences and Engineering Research Council of Canada.
 

 
\clearpage
 
\figcaption[grundahl_fig1.eps]
      {
       The ($u-y)_0, V$) color--magnitude
       diagram of M13, on which 10, 12, 14 Gyr isochrones (for [Fe/H] $=-1.61$ 
       and [$\alpha$/Fe] $=0.3$), as well as local subdwarfs having \hipparcos
       parallaxes ({\it filled circles}), have been superposed.  In the order
       of increasing $M_V$, the subdwarfs are: HD 34328, HD 64090, HD 188510,
       and HD 25329.  (Adding in HD 19945 and HD 29907 ({\it open squares}), 
       which have [Fe/H] $=-1.91$ and [Fe/H] $=-1.63$, respectively
       results in the same distance modulus).  The isochrone/ZAHB colors
       were adjusted by $\delta(u-y) = 0.05$ mag
       in order to achieve coincidence of the predicted and observed lower main sequences. 
       We attribute the failure of the models to match the RGB to problems with the Kurucz (1992) color
       transformations: R.~Bell (1998, private communication) has obtained
       much better agreement of the same isochrones with the M13 C-M
       diagram (in the vicinity of the turnoff) using his color-$T_{\rm eff}$
       relations.
       \label{grundahl_fig1}
      }
 
\figcaption[grundahl_fig2.eps]
      {The ($(b - y)_0, c_0$ ) diagram of M13, with field 
       stars having $-1.9 < [Fe/H] < -1.4$ from \cite{schuster89b} overplotted 
       as {\it filled circles}. Note the large scatter in $c_0$ on the cluster 
       RGB (but not among the few field giants that are present in the plot) at 
       any given $(b - y)_0$. \label{grundahl_fig2} 
      }
 
\clearpage
 
\epsscale{0.67}
\plotone{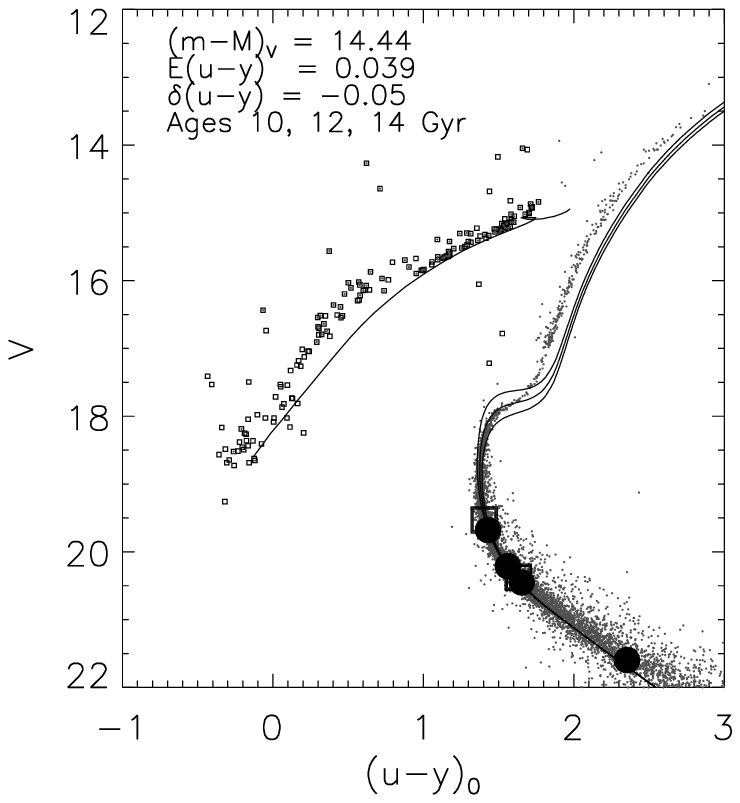}
 
\clearpage
 
\plotone{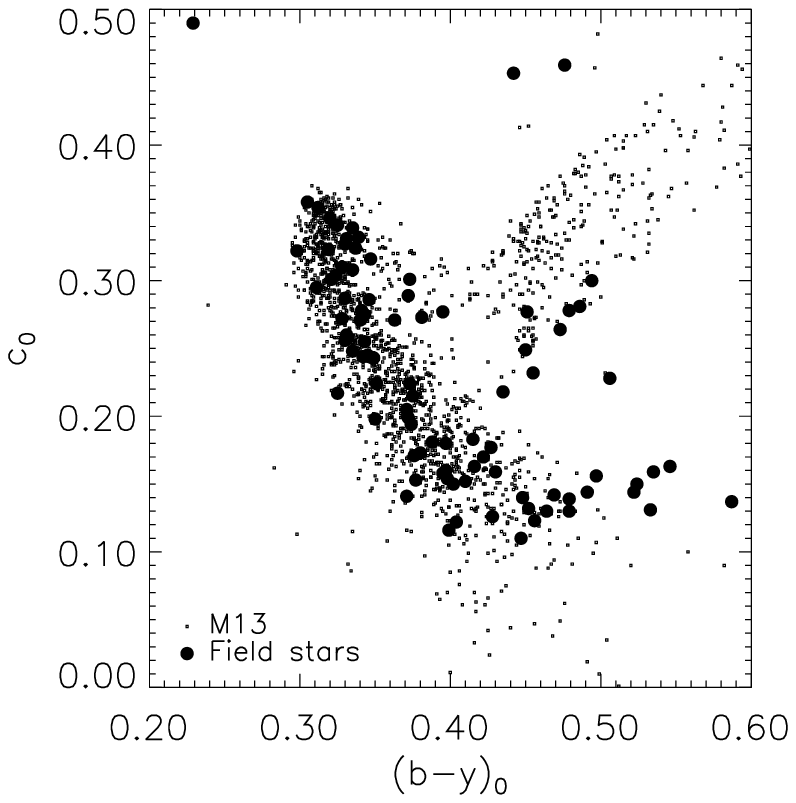}
 
\end{document}